\documentclass[journal]{IEEEtran}

\usepackage{latexsym, graphicx}
\usepackage{epsfig}
\usepackage{amsfonts, amssymb, amsthm}
\usepackage{float}
\usepackage{fancyhdr}
\usepackage{graphics}
\usepackage{stfloats}
\usepackage{multirow}
\usepackage{multicol}
\usepackage{amsmath}
\usepackage{bm}
\usepackage{array}
\usepackage{cite}
\usepackage{tabularx}
\usepackage{enumerate}
\usepackage{times}
\usepackage{algorithm}
\usepackage{algorithmic}
\usepackage{threeparttable}
\usepackage{lipsum}
\usepackage{indentfirst}
\usepackage{color}
\usepackage[caption=false]{subfig}
\usepackage{steinmetz}
\usepackage{makecell}
\usepackage{bbm}
\usepackage{booktabs}

\begin{document}

\title{
Power Margin Ratio - A Large-Signal System Strength Metric for Inverter-Based Resources-Dominated Power Systems
}

\author{Zitian~Qiu,~\IEEEmembership{Member,~IEEE,}
~Yunjie~Gu,~\IEEEmembership{Senior Member,~IEEE,}

\thanks{Zitian Qiu and Yunjie Gu are with the Department of Electrical and Electronic Engineering, Imperial College London, London, SW7 2BX, UK (E-mail: z.qiu@imperial.ac.uk,
yunjie.gu@imperial.ac.uk).}
}

\maketitle

\begin{abstract}
As the growing penetration of inverter-based resources (IBRs) in modern power systems, the system strength is decreasing. Due to the inherent difference in short-circuit capacity contributions of synchronous generators and inverters, the short-circuit ratio is not a one-size-fit-all metric to assess the system strength. Following the distinct dynamic behavior of the IBR in small- and large-signal disturbance, the system strength is separated accordingly. To address the large-signal system strength assessment, a control type-dependent metric, Power Margin Ratio (PMR), is proposed in this paper. PMR is defined as the ratio between the maximum power that can be injected to the system without causing any instability and the nominal power of the IBR. It can be obtained via power flow calculation with a modified algorithm. The theoretical foundation of PMR is established from the viewpoint of dynamical systems. PMR is identical to SCR for the single-plant-infinite-bus system, while presents advancement for multi-infeed power systems. Comprehensive case studies and discussions have validated that PMR reveals the large-signal system strength from a static perspective.
\end{abstract}


\IEEEpeerreviewmaketitle

\section{Introduction}

\IEEEPARstart{T}{HE} power system is transforming from a system dominated by synchronous
generators (SGs) to the one where inverter-based resources (IBRs) become dominant \cite{blaabjerg2004power}. Particularly, in the GB system, thermal power plants with synchronous generators are decommissioned in favour of IBRs in the drive to meet the UK’s net-zero targets \cite{gu2022power}.

As the increasing penetration of IBRs in modern power systems,  challenges in terms of operation performance and stability have been spotted to light. One of the issues that has drawn growing attention, especially in the GB, is the reduction of the system strength \cite{Henderson2024Grid}. System strength (or equivalently grid strength) encompasses a broad range of aspects and their implications on power system operability. It is defined as the ability of power system equipment to operate in a stable manner and for the system as a whole to recover intact from major disturbances \cite{badrzadeh2021systemstrength}. Typically, system strength is used, on the one hand, during connection studies to assess whether the system has enough ability to absorb the generation from the newly planned device, and on the other hand to assess how far a generator is from the limitation of power transfer.

In conventional power systems that are dominated by SGs, the short-circuit capacity (SCC), or short-circuit level (SCL), short-circuit ratio (SCR), is the standard metric of grid strength for evaluating the ability to connect a new device at a specific location. The SCL can be expressed as the SCR, the ratio between the current provided during a three-phase-to-ground fault to the nominal current \cite{Henderson2024Grid}. Typically, the short-circuit current contribution (also known as fault current contribution) of an SG can reach 5-7 p.u. \cite{johnson2017rfg} because of their low impedance and ability to withstand currents well above normal for short periods without a large and damaging temperature increase. Hence, the SCR defined based on SCL is consistent with the SCR defined by the Thevenin equivalent impedance, since the short-circuit current of the synchronous generator is entirely determined by its impedance, without any current limiters \cite{Zheng2023the}. 

However, due to the overcurrent limiters of the IBRs to protect their IGBTs, the SCL is no longer a quasi-linear output characteristic as SGs. This lead to the SCR based on the short-circuit capacity definition and the SCR based on the Thevenin equivalent impedance definition being fundamentally different for the IBRs. As pointed out in \cite{Zhu2024TPS}, it is a need to separate the system strength to small- and large-signal perspective since SCR is not a good overall indicator for all aspects of grid strength for an IBR-dominated power system. The small-signal system strength that considers the dynamic and possibly oscillatory behaviour of system voltage in response to small perturbations around the operating point. While the large-signal system strength concerns the fault recovery of a bus following large disturbances such as a local three-phase fault or a deep voltage dip caused by a remote fault. 

To assess the large-signal system strength, one of the most straightforward approach is to acquire the SCL of the concerned bus. The electromagnetic transient (EMT) simulations can accurately analyze the fault current and 
voltage during the fault and steady states. However, dynamic behavior studies require full knowledge of the system, including the detailed design of the IBR, which is usually not available to system operators \cite{Zhu2024TPS}. Even though such white-box models are accessible, EMT simulations are time consuming.

In practice, the large-signal system strength of IBR-dominated system is still using SCR as the metric. The most common approach is to ignore the contribution of IBRs from SCR calculations \cite{IEC60909-4_2021,Henderson2024Grid}. To incorporate the impact of IBRs in SCR calculation, in academic studies, some variants of SCR have been investigated lately. The fundamental idea is to manage to form an equivalent IBR connected to the concerned bus. Composite short-circuit ratio (CSCR) is proposed in \cite{GE_MISO_2014}, where IBRs connected to a bus through different lines' impedance are aggregated as an equivalent one. Similarly, weighted short-circuit ratio (WSCR) proposed in \cite{Zhang2014PESGM} assesses the short-circuit capacity separately, and then the WSCR is formed by weighting the values according to the power of the IBRs. However, both CSCR and WSCR omit the potential for interaction among the newly connected IBRs. Equivalent circuit-based short-circuit ratio (ESCR) is further developed in \cite{CIGRE_2016_WindWeakAC} for assessing the interaction among IBRs by applying an interaction factor, which is defined by introducing a small voltage change at one point of interconnection (POI) and observing the change of bus voltage at another POI. Although the IBR contribution is involved in those metrics, the different control modes of IBRs are not explored. For example, the grid-forming (GFM) inverter can fix the voltage and angle of the POI and is usually considered as a positive role which can increase the system strength \cite{Matevosyan2019Mag}. Unfortunately, this is not well-addressed in the SCR-based definitions.

On the other hand, large-signal stability criteria has the potential to be a metric for system strength assessment since it also concerns whether the system can recover from a large-signal disturbance. It relies on methods to quantify the size of the region of attraction (ROA) of the post-fault stable operating point\cite{Chiangbook}. If the system state remains inside the ROA when the fault is cleared, the post-fault system can maintain stable. Hence, larger ROA size implies a stronger the grid in terms of large-signal. However, the limitations of this category lies in two aspects, First, a detailed state space model is required while is generally unavailable from the vendors. Second, although the equal area criteria (EAC) \cite{Li2024TPS} and topological characterization \cite{Zhang2025arxiv} are effective for a single plant system and second-order dynamical systems, they face challenges to be implemented in a high-order multi-IBR system. An open question is to effectively and less conservatively estimate the ROA of large-scale power systems \cite{Qiu2023TPS}. 

Another issues is that short-circuit current contributions of transmission networks have not commonly included loads other than constant impedance loads \cite{Haddadi2023TPD}. It is reported that due to the comparable short-circuit current contribution for IBRs and loads, the neglection of loads may bring ill-informed insights into the system strength.

To address the above problems, a large-signal system strength metric termed power margin ratio (PMR) is proposed. By exploiting the different control type of IBRs, the PMR can be obtained via power flow calculation with a modified algorithm. Similar to SCR, a larger PMR value indicates the system is stronger. The contributions of this paper are three-folds.
\begin{itemize}
    \item A control type-dependent metric, PMR, is proposed for large-signal system strength assessment of IBR-dominated power systems. PMR addresses the problems that existing metrics capture only local interactions between IBRs or simply ignore the contributions of IBRs.
    \item PMR is a power flow calculation-based metric. Therefore, PMR provides a technology-neutral definition of system strength that is naturally suitable for system-wide, multi-infeed, source-load coexisted power systems. Moreover, a detailed state-space model from the vendor is not needed.
    \item The theoretical foundation of PMR is established, i.e., small PMR implies a small size of ROA. Hence, the system is prone to lose stability subjected to a large-signal disturbance.
\end{itemize}

\section{Distinction between small- and large-signal problems}\label{Section 2}

We now use simulation results to visualize that the system suffers large-signal disturbance does not manifest typical small-signal instability characteristic.

\begin{figure}[!t]
  \centering
  \includegraphics[width=3.4 in]{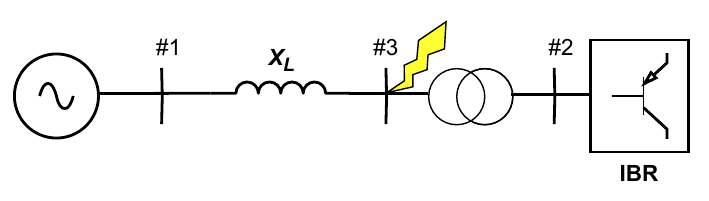}\\
  \caption{Single line diagram of the single-plant-infinite-bus system}\label{fig: SPIB system}
\end{figure}

Consider the single-plant-infinite-bus (SPIB) system shown in Fig. \ref{fig: SPIB system}, where all physical quantities are in per unit. Let the line impedance $X_L=1/1.12$ with resistance ignored, and the nominal power of the IBR $P_{nom}=1$. Here, we assume that the IBR is also injecting $1$ p.u. active power into the grid. The IBR is a gird-following (GFL) inverter with PV control. A three-phase short-circuit is triggered at Bus $3$ at $50$ s and lasts for $4$ cycles. We set high fault resistance to model the small disturbances and small fault resistance for large disturbances.
\begin{figure} [H]
\centering
\begin{minipage}{0.48\linewidth}
\centerline{\includegraphics[width=5cm]{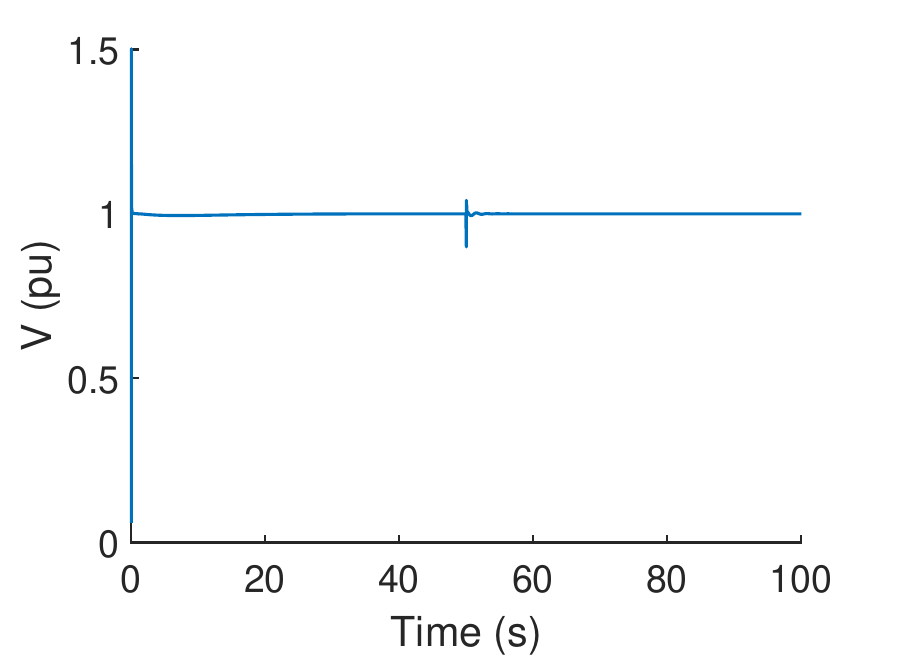}}
\centering
\subfloat{(a)}
\end{minipage}
\hfill
\begin{minipage}{0.48\linewidth}
\centerline{\includegraphics[width=5cm]{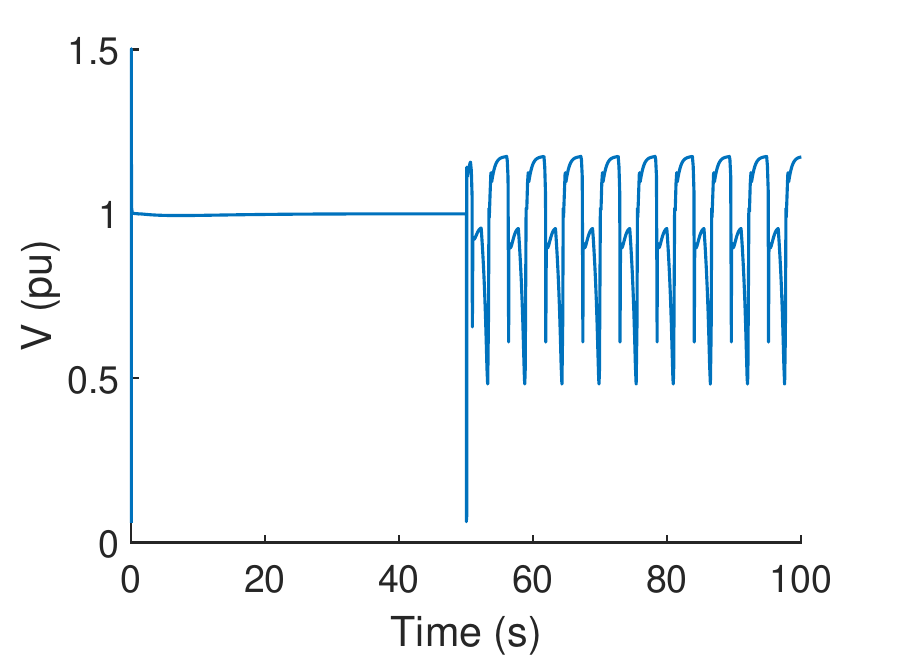}}
\centering
\subfloat{(b)}
\end{minipage}
\vfill
\caption{Voltage at Bus $\#3$ based on EMT simulations of the SPIB system under (a) small disturbance and (b) large disturbance.} \label{fig:Small_large_compare}
\end{figure}

The EMT simulation results are presented in Fig. \ref{fig:Small_large_compare}. It is clear in Fig. \ref{fig:Small_large_compare} (a) that when the system suffers a small disturbance, the voltage drops to $0.9$ p.u. The voltage of the IBR bus can recover after the fault, implying that the system is stable in the small signal sense. On the contrary, when the voltage drops significantly during the fault, the voltage of the IBR bus cannot recover after the fault is cleared as shown in Fig. \ref{fig:Small_large_compare} (b). It leads to large-signal instability of the system. Moreover, the large-signal instability does not exhibit oscillations with increasing amplitude, which is different from the typical small-signal instability mechanism. Hence, we can conclude that such instability is not a small-signal problem. As a result, it is inadequate to use small-signal metrics for large-signal stability assessment.

\section{Power margin ratio}
This section starts by briefly reviewing the calculation of the SCR, followed by a detailed illustration of the proposed PMR. Our understanding and comments on PMR are summarized at the end of this section.
\subsection{Short-circuit ratio}
Since SCR has been widely used for system strength assessment of SG-dominated power systems, it is important to review its concept and principle and its limitations to be adopted in IBR-dominated power systems.
\begin{figure}[!t]
  \centering
  \includegraphics[width=3.4 in]{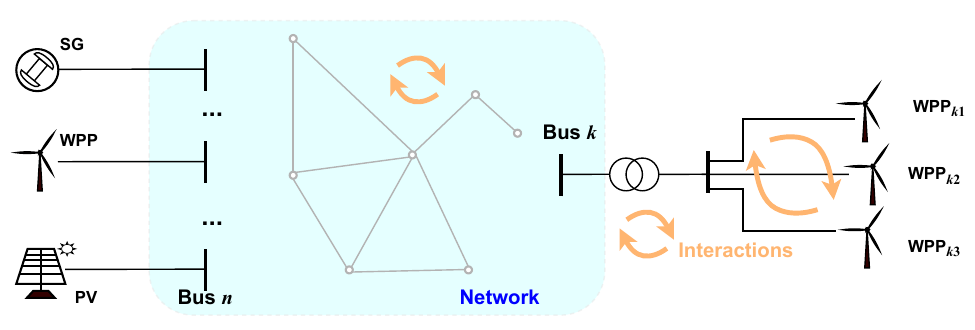}\\
  \caption{A typical structure of an IBR-dominated power system.}\label{fig: General system}
\end{figure}

\begin{figure}[!t]
  \centering
  \includegraphics[width=3.0 in]{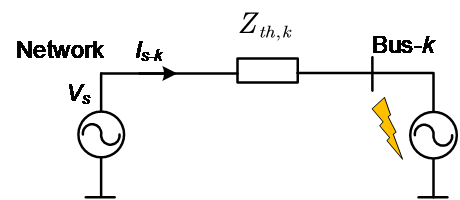}\\
  \caption{Circuit representation of a plant connected to an infinite bus.}\label{fig: Thevenin equivalence}
\end{figure}

Fig. \ref{fig: General system} presents a typical structure and components of an IBR-dominated power system, where wind power plants (WPPs) and photovoltaic (PV) plants are inverter interfaced power generation units. Regardless of the generation-side control to achieve power converting, we focus on the interaction of the grid-side converter with the grid. When viewed from a single bus, e.g., bus $k$, a model equivalence can be made on the rest of the system to create an equivalent ideal voltage source in series with a Thevenin impedance, as shown in Fig. \ref{fig: Thevenin equivalence}. $V_s$ is the equivalent Thevenin voltage; $Z_{th,k}$ is the Thevenin impedance seen from bus $k$; and $I_{s,k}$ is the current flow from the source.

When a three-phase-to-ground short-circuit happens at bus $k$,the power flows from the source to bus $k$ is defined as the short-circuit capacity, or short-circuit level at bus $k$, and is expressed as
\begin{equation}
 SCC_k=\frac{|V_s|^2}{|Z_{th,k}|}.
\end{equation}
SCR is then defined as the ratio between $SCC_k$ and the nominal power $P_{nom,k}$ of the device connected to bus$k$, i.e.,
\begin{equation}\label{Eqn: SCR pu}
 SCR_k=\frac{SCC_k}{P_{nom,k}}=\frac{|V_s|^2}{|Z_{th,k}|P_{nom,k}}.
\end{equation}

In a conventional power system where only SG exists, the SCR defined based on short-circuit capacity is completely consistent with the SCR defined by the Thevenin equivalent impedance, since the short-circuit current of the synchronous generator is entirely determined by its impedance, without any current limiters \cite{Zheng2023the}. However, due to the overcurrent limiters of the IBRs, the short-circuit capacity is no longer a quasi-linear output characteristic. Therefore, the SCR based on the short-circuit capacity definition and the SCR based on the Thevenin equivalent impedance definition are fundamentally different for the IBRs. As revealed in Section \ref{Section 2}, there exists drastic differences between small- and large-signal dynamics of the IBR-to-grid system. To address the limitations of SCR-based metrics for (large-signal) grid strength evaluation of IBR-dominated systems, we propose an alternative indicator named \textit{Power Margin Ratio}.

\subsection{Power margin ratio}
The basic idea behind using the PMR to assess the system's strength is to determine the maximum power a bus can handle while still maintaining a static operating point. Compared to the short-circuit current calculation which requires the "source + impedance" combination of the plant, power itself is a more intuitive and direct information for grid strength evaluation since it can be obtained by power flow calculation.

Given a power system as shown in Fig. \ref{fig: General system}, the PMR of bus $k$ is calculated as
\begin{equation}\label{Eqn: PMR}
    PMR_k=\frac{P_{max,k}}{P_{IBR}},
\end{equation}
where $P_{max,k}$ is the maximum power the system can absorb from this POI and $P_{IBR}$ is the nominal power of the connected IBR. $P_{max,k}$ can be acquired from power flow calculation by increasing the output power at the corresponding POI until reaching a non-converged result.

When conducting power flow calculations, we specify the node types for different generation plants as in Table \ref{tab: Node_type}. For an inverter based on GFL with $PQ$ control, we define it as a GFL (control). For the one adopts GFL with $PV$ control, we define it as the grid supporting, and GFM preserve its fundamental definition. It should be noted that in steady state power flow calculation, the GFM can be treated as a $PV$ bus. However, in PMR calculation, the power flow refers to the one within a short transient period. As a result, the GFM is assumed to be a stiff $V\theta$ node, where the power angle aligns with the pre-fault value. 
\begin{table}[!t]
    \centering
    \caption{Node types adopted in PMR calculation}
    \begin{tabular}{c|c}
    \hline
     Plant type    &  Node type\\
     \hline
     GFL    & $PQ$\\
     Grid supporting  & $PV$\\
     GFM  & $V\theta$\\
     Infinite bus & Slack\\
     \hline
    \end{tabular}
    \label{tab: Node_type}
\end{table}

Similar to the SCR, a larger PMR implies a stronger grid, i.e., the system is more likely to operate in a stable manner and for the system as a whole to recover intact from major disturbances \cite{badrzadeh2021systemstrength}.

\subsection{Relationship between PMR and SCR}
For the SPIB system shown in Fig. \ref{fig: SPIB system}, the PMR is identical to the conventional SCR. The reason is that the POI is the only point where an IBR can be integrated. The interaction can only take place between the IBR and the infinite bus, where the infinite bus is a stiff voltage source with constant voltage magnitude and frequency. Hence, the Thevenin impedance $X_L$ entirely captures the grid strength in the per-unit system, as given in (\ref{Eqn: SCR pu}). On the other hand, the active power that can be transferred between the IBR and the infinite bus is 
\begin{equation}\label{Eqn: Power transfer equation}
    P(\delta)=\frac{V_sV_i}{X_L}\sin(\delta),
\end{equation}
where $V_s$ and $V_i$ are voltages of the infinite bus and IBR; $\delta$ is the angular difference between two nodes. Therefore, the maximum power delivered is $P_{\max}=1/X_L$ in p.u. Substituting $P_{\max}$ into (\ref{Eqn: PMR}) and comparing it with (\ref{Eqn: SCR pu}) shows that PMR and SCR are identical for the SPIB system.

However, it becomes different in multi-IBR systems. In general, the contributions of IBRs to the grid strength are ignored when calculating the SCR of IBR-dominated power systems. From the power flow viewpoint, the IBRs are assumed to inject zero power into the grid, while the PMR involves the actual and potential power output with respect to the type of IBRs.

\subsection{Discussion on PMR}
PMR provides a technology-neutral definition of system strength that is naturally suitable for multi-infeed, large-signal system analysis and is simple to calculate. Moreover, GFL, grid supporting, and GFM inverters are classified as different nodes in power flow calculation, leading to a type-dependent large-signal system strength metric. As such, for instance, one can clearly reveal the benefit of integrating GFM inverters to improve the grid strength. Precisely, a larger PMR can be obtained if a GFL inverter is replaced by a GFM inverter. Since PMR requires power flow calculation, the interaction among different buses is naturally involved. It is worth noting that to calculate the PMR, one may have to modify the conventional power flow algorithm to include multiple $V\theta$ buses in the case where GFM inverters exist.

The PMR is defined with the nominal power of the IBR in (\ref{Eqn: PMR}). This is actually the "worst case" definition of the system strength. For a specific operating scenario, it is fair to calculate the PMR with the IBR's actual output power.

PMR cannot reveal exactly how the transient power flows, and it is not a rigorous certificate for the transient stability of power systems. It cannot assert whether the system is stable or not after a disturbance. In fact, a precise fault recovery is closely regulated by the control implementation and interactions of different control loops, which are beyond the scope of this paper since we focus on an indicator to evaluate the strength of a system node rather than transient stability analysis.

\section{Theoretical background of PMR}
In this section, we impose the theoretical foundation underlying PMR. From the viewpoint of dynamical systems, we point out that a small PMR reflects a small size of the ROA of a given equilibrium, implying that the system is more likely to lose stability if subjected to a large-signal disturbance. Instead of fully capturing the exact ROA, in this paper, we use the minimum distance between SEP and UEP as an estimation indicator to quantify the ROA size.

\subsection{A graphical interpretation}\label{Section: SPIB theory}
\begin{figure}[!t]
  \centering
  \includegraphics[width=3.4 in]{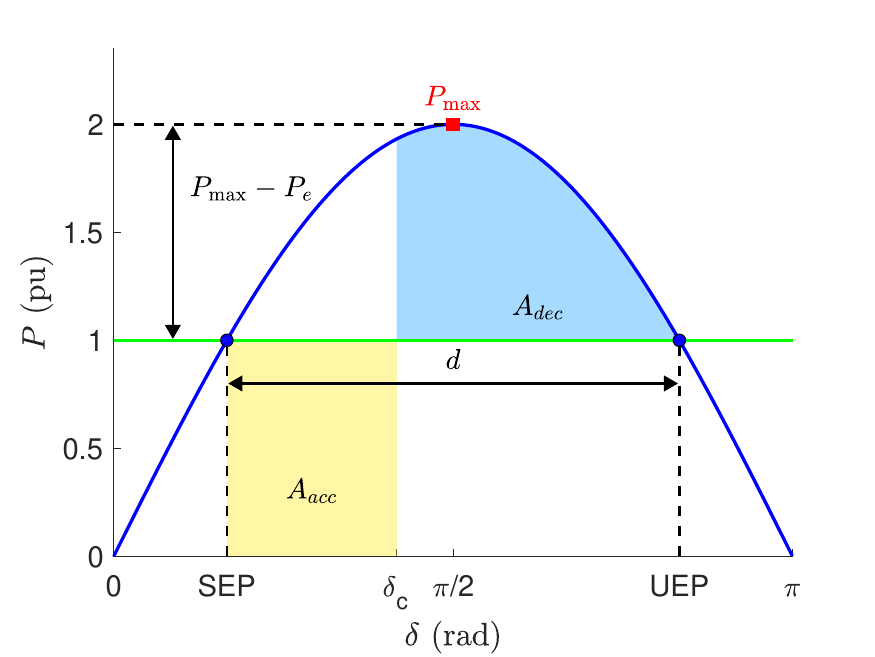}\\
  \caption{Illustration of $P-\delta$ curve of the single-plant-infinite-bus system.}\label{fig: SPIB Theory}
\end{figure}

We first point out the basic idea of PMR by using a typical single-machine-infinite-bus (SPIB) system and the equal area criterion (EAC). Consider a system with ignored line resistance, let $\bar{V_g}=V_g\angle0$ and $\bar{V}=V\angle\delta$ denote the voltage phasors of the infinite bus and the IBR, respectively. The power transferred from the IBR to the infinite bus is regulated by (\ref{Eqn: Power transfer equation}). Given bus voltages, the power delivered from the IBR is a function of the angular difference. The typical $P-\delta$ curve is shown in Fig. \ref{fig: SPIB Theory}, where the green line, denoted as $P_e$, is the nominal power (or actual output power) of the IBR. $P_e$ intersects $P(\delta)$ at two equilibrium points (EPs). The one that lies in $\delta\in[0,\pi/2]$ is a stable EP (SEP) since the real parts of eigenvalues of the linearized system are all negative, and the other one is an unstable EP (UEP).

Theoretically, given a voltage level and a line impedance, the maximum power admitted by the line without losing stability is $P_{max}=V_g V/X$, shown as the red dot in Fig. \ref{fig: SPIB Theory}. Hence, PMR can be interpreted as the distance between the maximum power point $P_{max}$ and the nominal power $P_e$. Recall that based on the EAC method, if the deceleration area is greater than the acceleration area, i.e., $|A_{dec}|>|A_{acc}|$, the system is stable after a disturbance. As such, given a fixed disturbance-cleared state $\delta_c$, a smaller PMR indicates the ''green'' line is moving upward towards $P_{\max}$, resulting in a decrease of deceleration area. The reduction of $A_{dec}$ implies the system is more likely to lose stability after being disturbed.

From the dynamical system viewpoint, the decrease of $A_{dec}$ is a manifestation of the shrinkage in the stability region (or the region of attraction, ROA) of the system's SEP \cite{Chiang1995Proc}. Based on the energy function method, the stability boundary of the SEP is composed of the stable manifold of UEPs \cite{Chiangbook}. Note that as PMR decreases, the distance $d$ between SEP and UEP also get smaller. It is reasonable to use the distance between SEP and UEP as an indicator to unrigorously evaluate the size of the stability region. Hence, we devote to explore a relationship
\begin{align}\label{Eqn: PMR and ROA}
    PMR\downarrow \Rightarrow d\downarrow \Rightarrow ROA\downarrow \Rightarrow \text{Weak grid}.
\end{align}

\subsection{The relationship between PMR and stability region}\label{Section PMR ROA}

Assume a power system ha has $n_{PV}$ $PV$ buses and $n_{PQ}$ $PQ$ buses, and let $n_b=n_{PV}+n_{PQ}$. Define $\bm{\theta}=[\bm{\theta}_{PV};\bm{\theta}_{PQ}]\in\mathbbm{R}^{n_b}$ where $\bm{\theta}_{PV}$ and $\bm{\theta}_{PQ}$ are composed of voltage angle of $PV$ and $PQ$ buses, respectively. $\bm{V}_{PQ}\in\mathbbm{R}^{n_{PQ}}$ collects bus voltage magnitudes of $PQ$ buses. The power flow equation of bus $i$ is
\begin{subequations}\label{Eqn: Power flow}
\begin{align}
    P_i(\bm{\theta},\bm{V})&= V_i\sum_{j=0}^{n_b}V_j(G_{ij}\cos\theta_{ij}+B_{ij}\sin\theta_{ij}),\\
    Q_i(\bm{\theta},\bm{V})&= V_i\sum_{j=0}^{n_b}V_j(G_{ij}\sin\theta_{ij}-B_{ij}\cos\theta_{ij}),
\end{align}
\end{subequations}
where $P_i$ and $Q_i$ are active and reactive power injected to bus $i$; $\theta_{ij}=\theta_i-\theta_j$; $G_{ij}$ and $B_{ij}$ are the conductance and susceptance of the line that connects to bus $i$ and $j$.

Assume Bus $c$ is the concerned bus, we introduce a new scalar $\lambda\leq0$ and rewrite the power equation as
\begin{equation}
    P_c(\lambda)=P_{\max}(1+\lambda).
\end{equation}
In fact, $|\lambda|_{max}$ can be interpreted as $P_{max}-P_e$ in Fig. \ref{fig: SPIB Theory}. Hence, the net active and reactive power at each bus are
\begin{subequations}
\begin{align}
    f_{P,i}(\bm{\bm{\theta},\bm{V}})&=P^0_i-P_i(\bm{\theta},\bm{V}),i\neq c,\\
    f_{P,i}(\bm{\bm{\theta},\bm{V},\lambda})&=P_{\max}(1+\lambda)-P_i(\bm{\theta},\bm{V}),i= c,\\
    f_{Q,i}(\bm{\bm{\theta},\bm{V}})&=Q^0_i-Q_i(\bm{\theta},\bm{V}),
\end{align}
\end{subequations}
where the superscript ``$0$" denotes the steady state value of active and reactive power at bus $i$.

Now, let $n=n_{PV}+2n_{PQ}$. We define a pseudo-dynamical system
\begin{equation}\label{Eqn: Pseudo system}
    \dot{\bm{x}}=\bm{f}(\bm{\theta},\bm{V},\lambda)=\bm{f}(\bm{x},\lambda),
\end{equation}
where $\bm{x}=[\bm{\theta}_{PV};\bm{\theta}_{PQ};\bm{V}_{PQ}]\in\mathbbm{R}^{n}$ is the state variable consisting $n_{PV}$ $PV$ bus voltage angles, $n_{PQ}$ $PQ$ bus voltage angles, and $n_{PQ}$ $PQ$ bus voltages; $\bm{f}(\cdot)=[\bm{f}_{P}(\bm{\bm{\theta},\bm{V},\lambda});\bm{f}_Q(\bm{\bm{\theta},\bm{V}})]\in\mathbbm{R}^n$ is the power flow equation induced $C^\infty$ vector field. The solutions of the power flow equations (\ref{Eqn: Power flow}) are the EPs of the system (\ref{Eqn: Pseudo system}). 

Assume that $\lambda_0\rightarrow0$ and the SEP is denoted as $\bm{x}_0$, linearize $\bm{f}(\bm{x},\lambda)$ at $(\bm{x}_0,\lambda_0)$ yields
\begin{equation}
    \dot{\bm{x}}=\bm{Ax},
\end{equation}
where $\bm{A}=\bm{J_xf}$ is the Jacobian matrix of $\bm{f(\cdot)}$ with respect to $\bm{x}$ at the point, i.e., $\bm{J_xf}=\partial\bm{f}/\partial\bm{x}$ at $(\bm{x}_0,\lambda_0)$.

Note that when $\lambda=0$, the rank of $\bm{A}$ is $n-1$ (also the minimum singular value is $0$), implying a fold bifurcation occurs \cite{Hiskens2021TPS}. Denote here the Jacobian matrix as $\bm{J}_0$. Let $\bm{\nu}$ be the kernel of $\bm{J}_0$, i.e.,
\begin{equation}\label{Eqn: Jnu}
    \bm{J}_0\bm{\nu}=\bm{0},
\end{equation}
and vector $\bm{\omega}$ such that
\begin{equation}\label{Eqn: omegaJ}
    \bm{\omega}^T\bm{J}_0=\bm{0}.
\end{equation}

Then, we nomalize $\bm{\nu}$ and $\bm{\omega}$ via $||v||=1$ and $\bm{\omega}^T\bm{\nu}=1$. Hence, a projection matrix $\bm{R}$ can be defined
\begin{equation}
    \bm{R}=\bm{I}-\bm{\nu}\bm{\omega}^T,
\end{equation}
where $\bm{I}$ is the $n\times n$ identity matrix.

Based on the center manifold theorem \cite{Kundur} (or Lyapunov-Schmidt reduction), the state space can be decomposed into a $1$-dimensional center subspace and an $n-1$-dimensional transverse subspace that consists of stable and unstable directions, i.e.,
\begin{equation}
    \mathbbm{R}^n=\text{ker} \bm{J}_0\oplus\text{Im}\bm{J}_0=\mathbbm{E}_C\oplus\mathbbm{E}_I,
\end{equation}
where $\text{Im}\bm{J}_0$ is the image space of $\bm{J}_0$, $\mathbbm{E}_C$ and $\mathbbm{E}_I$ denote the center subspace and the transversal subspace of $\mathbbm{E}_C$, respectively.

We next derive a nominal form of expression of the center manifold. Note that the movement of the state can be decomposed into or projected on $\mathbbm{E}_C$ and $\mathbbm{E}_I$. We hence define a coordinate transformation
\begin{subequations}\label{Eqn: Coordinate transformation}
    \begin{align}
        z&=\bm{\omega^T}(\bm{x}-\bm{x}_0), \\
        \bm{\psi}&=\bm{R}(\bm{x}-\bm{x}_0),
    \end{align}
\end{subequations}

From \ref{Eqn: Coordinate transformation} we can decompose $\bm{x}$ in terms of $z$ and $\bm{\psi}$ as
\begin{equation}
    \bm{x}=\bm{x}_0+z\bm{\nu}+\bm{\psi}.
\end{equation}

Then the original dynamical system (\ref{Eqn: Pseudo system}) is converted into $(z,\bm{\psi})$-coordinate as
\begin{subequations}
    \begin{align}
        \dot{z}&=\bm{\omega}^T\bm{f}(\bm{x}_0+z\bm{\nu}+\bm{\psi},\lambda),\label{Eqn: System with z coordinate}\\
        \dot{\bm{\psi}}&=\bm{Rf}(\bm{x}_0+z\bm{\nu}+\bm{\psi},\lambda).
    \end{align}
\end{subequations}

Next, we use Taylor expansion at $\bm{x}=\bm{x}_0$ and $\lambda=0$ to reformulate $\bm{f}(\cdot)$ as
\begin{align}\notag
    \bm{f}(\bm{x}_0+z\bm{\nu}+\bm{\psi},\lambda) =& \bm{J}_0(z\bm{\nu}+\bm{\psi})+\lambda P_{\max}\bm{e}_c\\
    &+\frac{1}{2}\Big[\bm{\nu}^T\bm{H}_i\bm{\nu}\Big]^n_{i=1} z^2+\bm{r}(z,\bm{\psi})+\mathcal{O}(z^3),\label{Eqn: Taylor of f}
\end{align}
where $\bm{e}_c$ is a vector with the $c$th element being $1$ and others being $0$; $[\bm{\nu}^T\bm{H}_i\bm{\nu}]^n_{i=1}=\Big[\bm{\nu}^T\bm{D}^2_{\bm{x}}\bm{f}_1\Big|_{\substack{\bm{x}=\bm{x}_0\\\lambda=0}}\bm{\nu},\cdots,\bm{\nu}^T\bm{D}^2_{\bm{x}}\bm{f}_n\Big|_{\substack{\bm{x}=\bm{x}_0\\\lambda=0}}\bm{\nu}\Big]^T$ forms an $n$-dimensional vector, and $\mathcal{O}(z^3)$ captures terms of order $z^3$.

We now estimate $\bm{r}(\bm{\psi})$. We are concerned with the EPs near the bifurcation point, hence setting $\dot{\bm{\psi}}=\bm{0}$ yields an algebraic constraint. Define 
\begin{equation}
\bm{F}(z,\bm{\psi},\lambda)=\bm{Rf}(\bm{x},\lambda), \bm{F}(0,\bm{0},0)=\bm{0}.
\end{equation}

The derivative of $\bm{F}$ with respective to $\bm{\psi}$ yields
\begin{equation}
\bm{D}_{\bm{\psi}}\bm{F}=\bm{RJ}_0,
\end{equation}
where $\bm{D}_{\bm{\psi}}\bm{F}$ is invertible. Based on the implicit function theorem,
\begin{equation}
\bm{\psi}=\bm{h}(z,\lambda), \bm{h}(0,0)=\bm{0},
\end{equation}
such that 
\begin{equation}
\bm{F}(z,\bm{h}(z,\lambda),\lambda)=\bm{0}.
\end{equation}

Now, we use Taylor expansion at $(0,0)$ to reformulate $\bm{\psi}=\bm{h}(z,\lambda)$ as
\begin{equation}
\bm{\psi}=\frac{\partial\bm{h}}{\partial z}\Big|_{(0,0)}z+\frac{\partial\bm{h}}{\partial\lambda}\Big|_{(0,0)}\lambda+\mathcal{O}(z^2,\lambda^2).
\end{equation}
Based on the implicit function theorem, the partial derivatives can be obtained via
\begin{subequations}
\begin{align*}
\frac{\partial\bm{h}}{\partial z}&=-(\bm{RJ}_0)^{-1}\frac{\partial\bm{F}}{\partial z},\\
\frac{\partial\bm{h}}{\partial \lambda}&=-(\bm{RJ}_0)^{-1}\frac{\partial\bm{F}}{\partial \lambda}.
\end{align*}
\end{subequations}
Hence, $\frac{\partial\bm{h}}{\partial z}\Big|_{(0,0)}=\bm{0}$ and $\frac{\partial\bm{h}}{\partial\lambda}\Big|_{(0,0)}=-(\bm{RJ}_0)^{-1}\bm{R}P_{\max}\bm{e}_c\neq 0$ in general. As a result, $\bm{\psi}=\bm{h}(z,\lambda)$ can be expressed as
\begin{equation}
\bm{\psi}=-(\bm{RJ}_0)^{-1}\bm{R}P_{\max}\bm{e}_c\lambda+\mathcal{O}(z^2,\lambda^2),
\end{equation}
i.e.,
\begin{equation}
\bm{\psi}\sim\mathcal{O}(z^2,\lambda).
\end{equation}
Therefore, we can estimate $\bm{r}(\bm{\psi})$ in (\ref{Eqn: Taylor of f}) as
\begin{equation}\label{Eqn: r term}
\bm{r}(z,\bm{\psi})\sim \gamma\lambda z+\mathcal{O}(\lambda z^2,\lambda^2z),
\end{equation}
where $\gamma$ is the coefficient of the second-order term $\lambda z$.

Substitute (\ref{Eqn: r term}) and (\ref{Eqn: Taylor of f}) into (\ref{Eqn: System with z coordinate}) and suppress all terms with order higher than $2$, we obtain
\begin{align*}
\dot{z}&=\bm{\omega}^T\bm{f}(\bm{x},\lambda)\\
&=\bm{\omega}^T\bm{J}_0(z\bm{\nu}+\bm{\psi})+\bm{\omega}^T\lambda P_{\max}\bm{e}_c\\
&\quad+\bm{\omega}^T*\frac{1}{2}\Big[\bm{\nu}^T\bm{H}_i\bm{\nu}\Big]^n_{i=1} z^2+\bm{\omega}^T\bm{e}_c\gamma\lambda z+\mathcal{O}(\lambda z^2,\lambda^2z)\\
&=\bm{\omega}^TP_{\max}\bm{e}_c\lambda+\bm{\omega}^T*\frac{1}{2}\Big[\bm{\nu}^T\bm{H}_i\bm{\nu}\Big]^n_{i=1} z^2+\bm{\omega}^T\bm{e}_c\gamma\lambda z.\\
&=\alpha\lambda+\beta z^2+\gamma\lambda z.
\end{align*}
We further apply a coordinate transformation to eliminate the interaction term $\lambda z$ via \cite{Kuznetsov}
\begin{equation*}
    z=Z-k\lambda,
\end{equation*}
where $k$ is a coefficient to be determined. With the new variable $Z$, we have
\begin{align*}
    \dot{Z}&=\alpha\lambda+\beta(Z-k\lambda)^2+\gamma\lambda(Z-k\lambda),\\
    &=\alpha\lambda+\beta Z^2-2k\beta\lambda Z+\gamma\lambda Z+\mathcal{O}(\lambda^2).\\
\end{align*}
Hence, we can solve $k=\frac{\gamma}{2\beta}$ to eliminate term $\lambda Z$. Note that this transformation is a local diffeomorphism with identity linear part, hence it preserves the fold singularity and only shifts the origin along direction $\bm{\nu}$ by $k\lambda$. As a result, by omitting the higher-order terms, the reduced 1-dimensional nominal form of 1-parameter fold bifurcation is obtained as
\begin{equation}\label{Eqn: 1-parameter bifurcation}
    \dot{Z}\approx\alpha\lambda+\beta Z^2,
\end{equation}
where
\begin{align*}
    \alpha&=\bm{\omega}^TP_{\max}\bm{e}_c,\\
    \beta&=\bm{\omega}^T*\frac{1}{2}\Big[\bm{\nu}^T\bm{H}_i\bm{\nu}\Big]^n_{i=1}.
\end{align*}

Therefore, the two EPs of (\ref{Eqn: 1-parameter bifurcation}) are
\begin{equation*}
    Z_{1,2}=\pm\sqrt{-\frac{\alpha}{\beta}\lambda},
\end{equation*}
and return to $z$ via
\begin{equation}
    z_{1,2}=\pm\sqrt{-\frac{\alpha}{\beta}\lambda}-k\lambda.
\end{equation}
The distance between two EPs projected on the center manifold is hence
\begin{equation}\label{Eqn: Estimated distance}
    d=2\sqrt{-\frac{\alpha}{\beta}\lambda},
\end{equation}
implying (\ref{Eqn: PMR and ROA}) with the relationship between PMR and $\lambda$.

\textit{Remark 1:} Since the vector field of the pseudo dynamical system (\ref{Eqn: Pseudo system}) is composed of power flow equations, the Jacobian matrix of system (\ref{Eqn: Pseudo system}) is the same as the one used in power flow calculation, i.e., the Newton-Raphson method. 

\textit{Remark 2:} Vectors $\bm{\nu}$ and $\bm{\omega}$ can also be obtained via singular value decomposition (SVD) of the Jacobian matrix $\bm{J}_0$ and fetching the vectors with respective to the smallest singular value (nearly 0). The merit of using a singular vector is that the singular vector is less numerically sensitive when the Jacobian is near singularity.

\textit{Remark 3:} The term $\bm{H}_i$ is indeed the Hessian matrix of $\bm{f}_i$ with respective to $\bm{x}$, which can be obtained along solving power flow equations.

\section{Case studies}
The simulations are carried out with Matlab/Simulink in an Intel CORE i5, 16 GB RAM laptop.

\subsection{Single-plant-infinite-bus system}
First, we adopt the SPIB system as a demonstration of the PMR. The system is given in Fig. \ref{fig: SPIB system}. Unless specified, the nominal power of the IBR is fixed at $1$ p.u., while the line's static power transfer limit (in terms of the line's impedance) is fixed at $X_L=1/1.13$. The equivalent impedance of the transformer is $0.05$ p.u. Based on (\ref{Eqn: PMR}), the calculated PMR of GFM, grid supporting, and GFL cases are $>5$, $1.05$, and $0.85$, respectively.

\begin{figure} [!t]
\centering
\begin{minipage}{0.48\linewidth}
\centerline{\includegraphics[width=5cm]{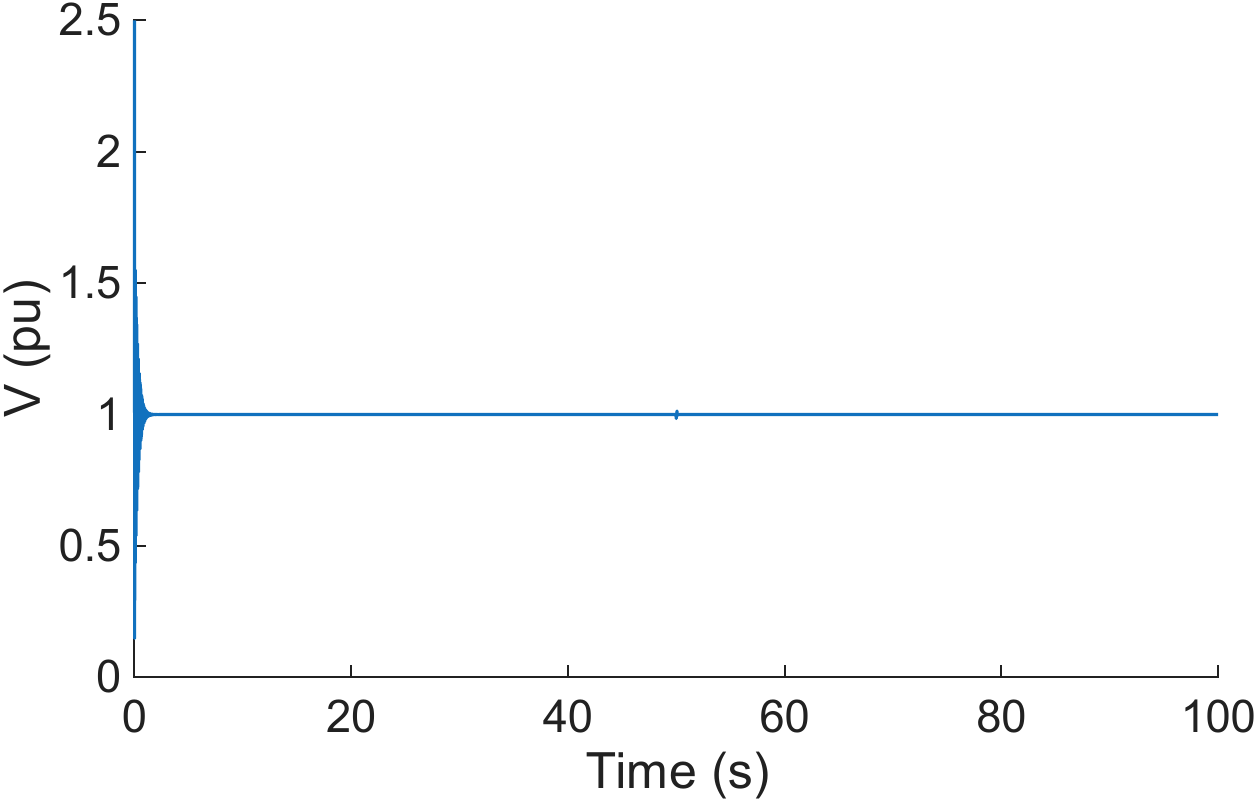}}
\centering
\subfloat{(a)}
\end{minipage}
\hfill
\begin{minipage}{0.48\linewidth}
\centerline{\includegraphics[width=5cm]{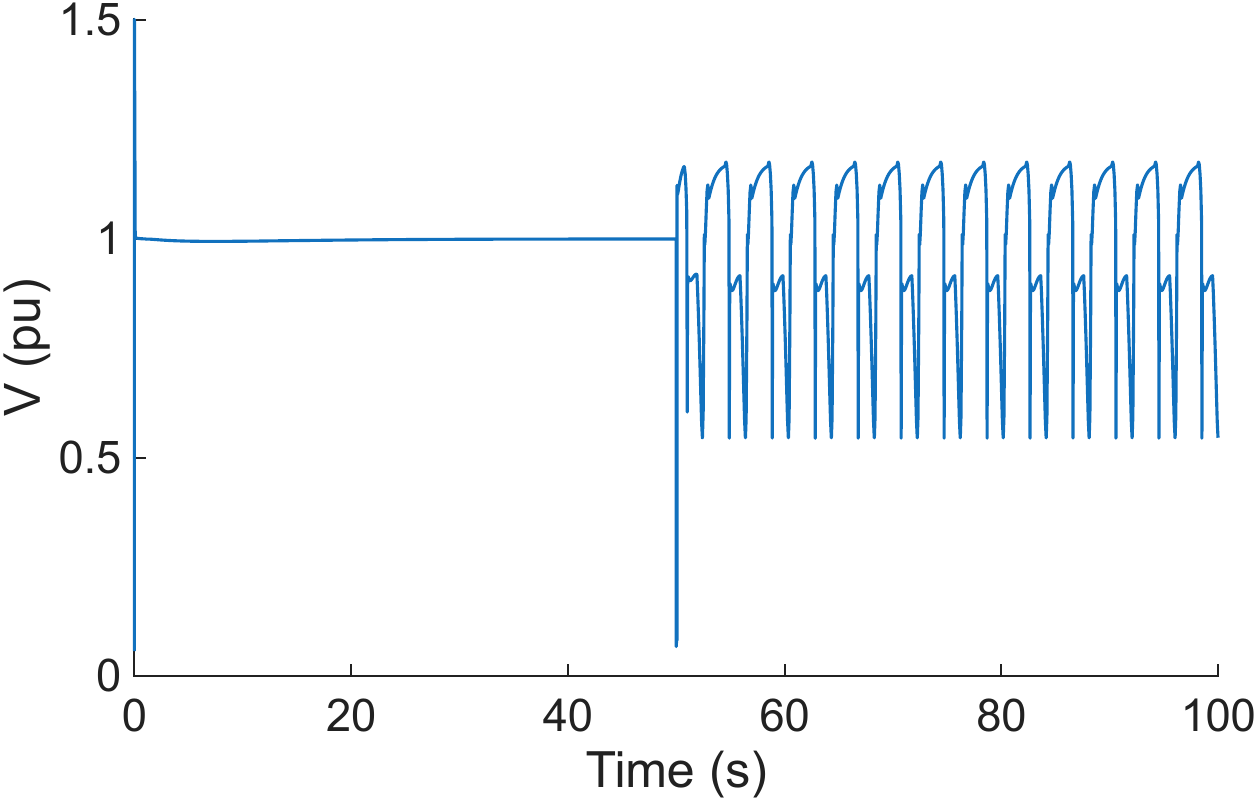}}
\centering
\subfloat{(b)}
\end{minipage}
\vfill
\begin{minipage}{0.48\linewidth}
\centerline{\includegraphics[width=5cm]{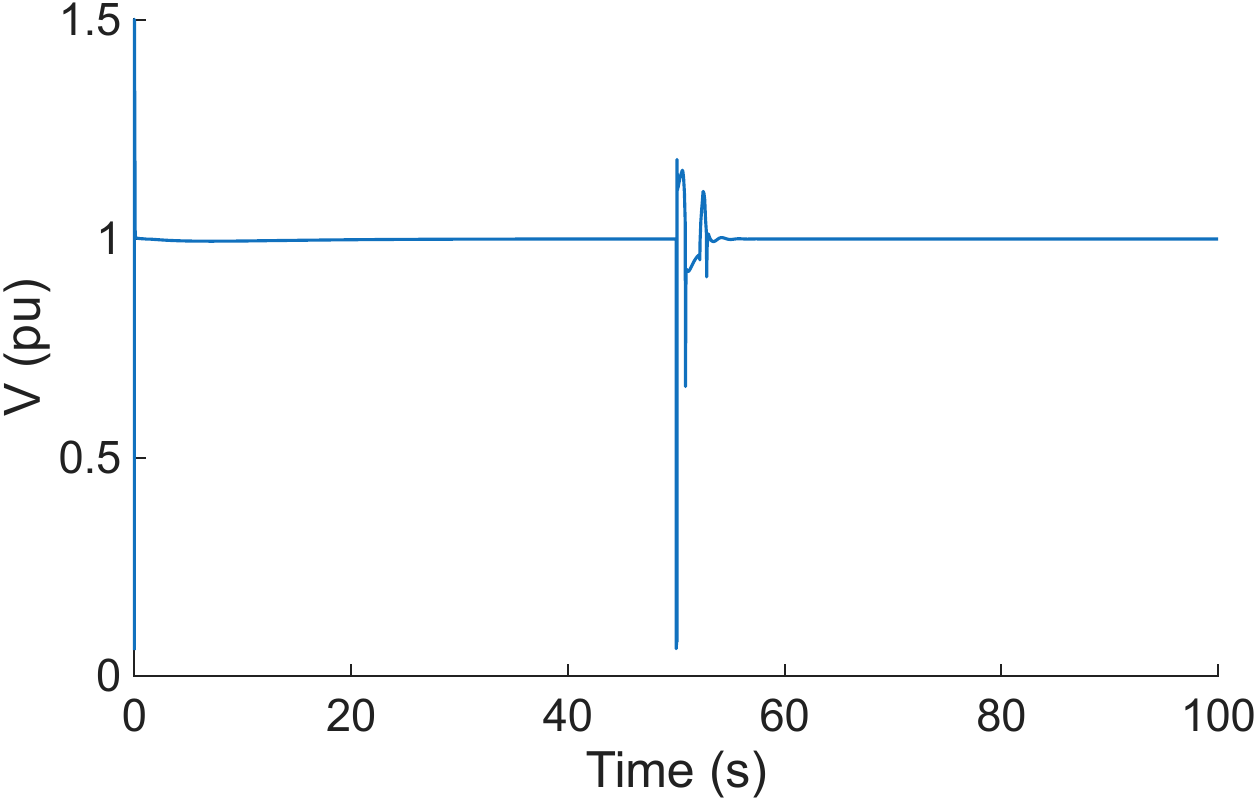}}
\centering
\subfloat{(c)}
\end{minipage}
\hfill
\begin{minipage}{0.48\linewidth}
\centerline{\includegraphics[width=5cm]{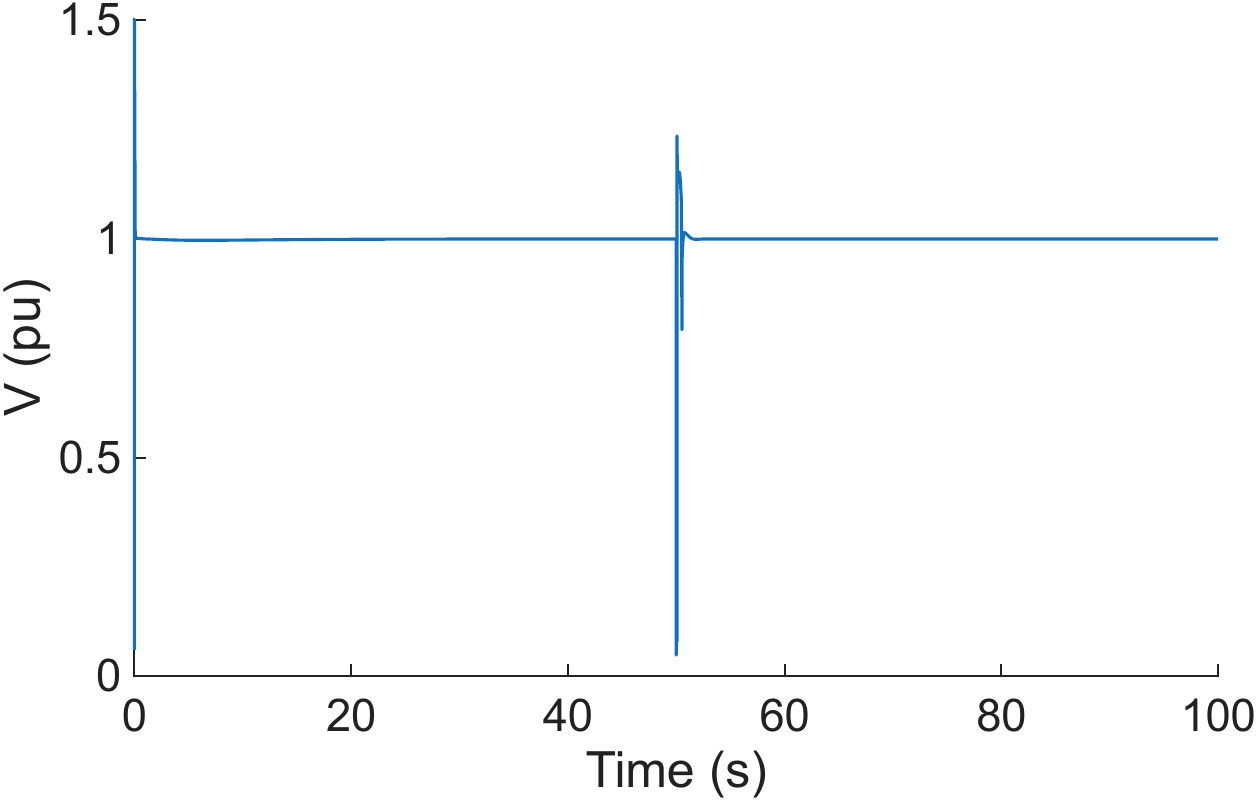}}
\centering
\subfloat{(d)}
\end{minipage}
\vfill
\begin{minipage}{0.48\linewidth}
\centerline{\includegraphics[width=5cm]{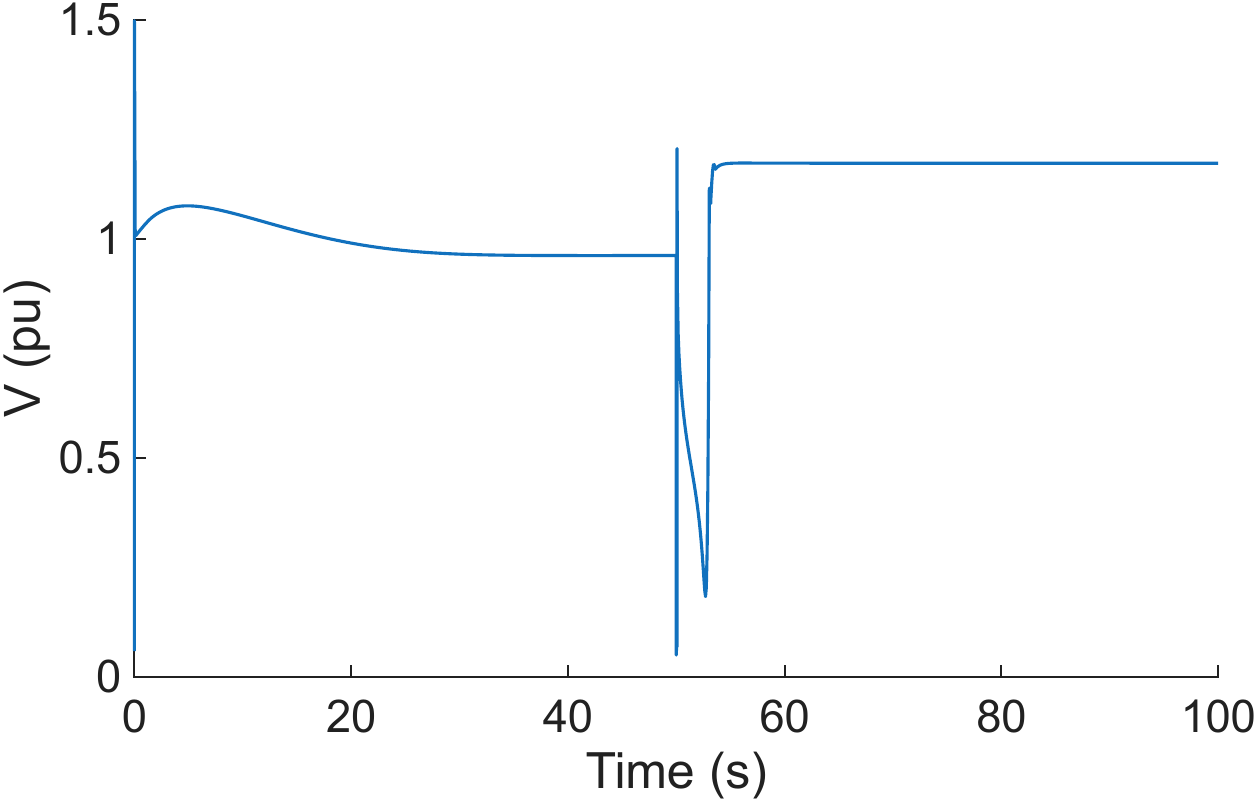}}
\centering
\subfloat{(e)}
\end{minipage}
\caption{Voltage of the IBR based on EMT simulations of the SPIB system under (a) GFM case where $P_{I}=1$, (b) grid supporting case where $P_{I}=1.05$, (c) grid supporting case where $P_{I}=1$, (d) grid supporting case where $P_I=0.8$, and (e) GFL case where $P_I=0.8$.} \label{fig:SPIB compare1}
\end{figure}

We carried out EMT simulations to verify the effectiveness of the PMR. A three-phase-to-ground short-circuit is applied at Bus $3$ and lasts for $4$ cycles. The voltages of the IBR based on EMT simulation results are presented in Fig. \ref{fig:SPIB compare1}. If the IBR is a GFM inverter, the system can withstand the fault, as shown in Fig. \ref{fig:SPIB compare1} (a). The GFM acts as a ``stiff" voltage source, facilitating the fault recovery precess. Next, we consider the IBR as a grid supporting inverter. When the output power of IBR is $1.05$, the system cannot recover from the fault, while reducing the power to $1$ p.u. can make the system withstand the fault. The results are illustrated in Fig. \ref{fig:SPIB compare1} (b) and (c). Further reducing the actual power to $P_I=0.8$ p.u. while keeping its nominal value as $P_{IBR}=1$ p.u., the system is stable after the fault is cleared as presented in Fig. \ref{fig:SPIB compare1} (d). Finally, let the IBR be a GFL inverter with nominal power being $P_{IBR}=1$ p.u. and actual output power being $P_I=0.8$. The system operates at a steady state in the pre-fault stage. However, it failed to recover from a fault due to the poor voltage regulation. These results confirm that the PMR can evaluate the system's dynamic behavior under large disturbances from a static viewpoint. Moreover, a higher PMR value implies a stronger grid. 

In practice, we can regard $PMR_c=1.05$ as the critical PMR for this fixed-topology system with adjustable IBR output power. Then, any PMR greater than $PMR_c$ ensures that the system can withstand the same short-circuit fault. Generally, given a fixed line impedance, $PMR>PMR_c$ may appear in the IBR with either smaller nominal power or reasonable nominal power but with less actual output power. The former can be directly deduced from Eqn. (\ref{Eqn: PMR}), and the latter can be obtained by using the actual output power of the IBR as the denominator in Eqn. (\ref{Eqn: PMR}). Fig. \ref{fig:SPIB compare1} (d) illustrates the voltage of IBR in a scenario where the nominal power is $P_{IBR}=1$ while the actual output power is $P_I=0.8$. It is evident that the system can withstand the fault. Since the actual power of IBR is smaller than $1$ p.u., the PMR becomes larger than $PMR_c$. As a result, reducing the normal operating output power of an IBR can enhance system strength of an SPIB system. Moreover, using the nominal power of an IBR generally induces the ``worst" system strength.

Note that the enhanced system strength as PMR increases can be visualized via Fig. \ref{fig: SPIB Theory} for this SPIB system. As discussed in Section \ref{Section: SPIB theory}, if we adopt the actual power as the denominator of (\ref{Eqn: PMR}), reducing the output power implies that the green line moves downward away from the static limit point. The distance $d$ increases. Hence, the deceleration area enlarges, while the acceleration area shrinks, which benefits the post-fault recovery process. 

\begin{figure}[!t]
  \centering
  \includegraphics[width=3.4 in]{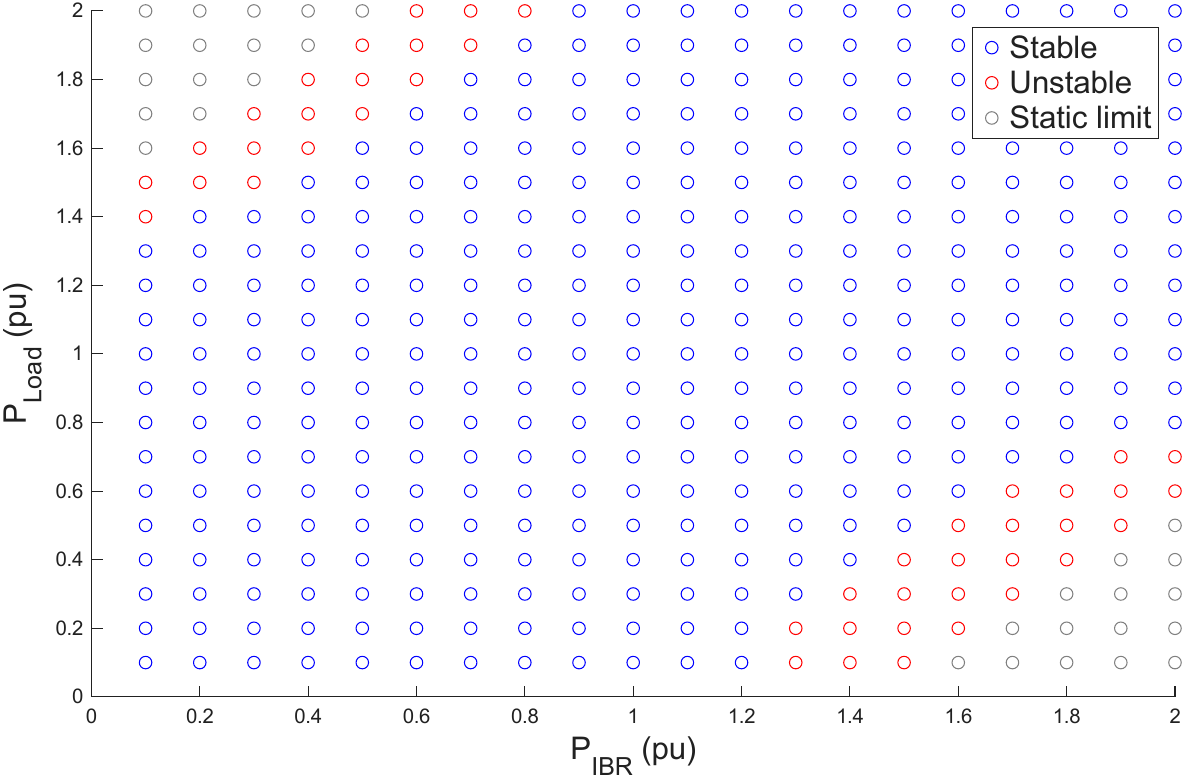}\\
  \caption{Stable and unstable cases of different combinations of IBR and CPL.}\label{fig: IBR_CPL}
\end{figure}

Now, we illustrate the impact of the load on PMR and system strength. Assume that a CPL is integrated at Bus $3$, where the CPL is modeled as a GFL with negative power reference. Moreover, following the switching logic of the electronic-interfaced loads given in \cite{epri2021compositeload}, we set the CPL operating at constant power mode if the voltage is above $0.5$ p.u., constant current mode if the voltage lies between $0.5$ and $0.2$ p.u., and cut off if the voltage below $0.2$ p.u.. We also assume that the voltage drop and increase process share the same segment curve. By EMT simulations, the stable and unstable cases of different combinations of IBR and CPL are visualized in Fig. \ref{fig: IBR_CPL}. The blue circles are stable cases, red ones are unstable combinations, and the gray circles indicates the net power exceed the line's power transfer limit.

Note that the line impedance is $X_L=1/1.5$, admitting $1.5$ p.u. in maximum. Let focus on the top right corner where $P_{IBR}>1.5$ and $|P_{Load}|>1.5$. If no local load exists, the IBR cannot generate $>1.5$ p.u. active power. However, since the local CPL absorbs power from the IBR, the net power seen from Bus $3$ becomes $<1.5$ p.u.. From PMR perspective, the IBR and CPL should be treated as a whole. By using the nominal power definition, the CPL increases PMR from $PMR=\frac{1.5}{2}<1$ to $PMR=\frac{1.5+|P_{Load}|}{2}>1.5$. On the other hand, by using the actual power definition, the nominator of PMR is fixed at $1.5$ while the denominator decreases from $P_{actual}=P_{IBR}>1.5$ to $P_{actual}=|P_{IBR}-|P_{Load}||\in [0,0.5]$, enhancing the PMR. We should point out that the unstable cases are caused by control interactions between IBR and CPL. If reducing the bandwidth of the PLL of the CPL, more unstable cases appears, narrowing the blue area towards the diagonal line.

\subsection{Two-IBR-infinite-bus system}
\begin{figure}[!t]
    \centering
    \includegraphics[width=0.9\linewidth]{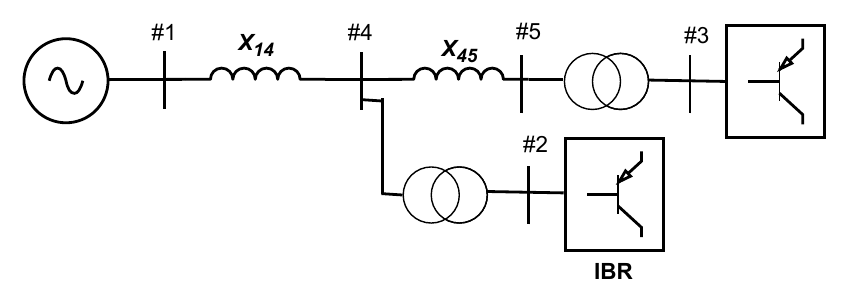}
    \caption{Single-line diagram and bus numbering of the two-IBR-infinite-bus system.}
    \label{fig: Two IBR system}
\end{figure}

We now consider a two-IBR-infinite-bus system. The single-line diagram and bus numbering of the system are shown in Fig. \ref{fig: Two IBR system}. For simplicity, we denote this system as the two-IBR system. All the faults triggered in this section are three-phase-to-ground short-circuit faults with low fault impedance. The fault lasts for $4$ cycles before cleared.

In Section \ref{Section: EMT high SCR}) and \ref{Section: EMT Low SCR}), we use EMT simulations to demonstrate that the SCR is not a proper metric for grid strength evaluation of IBR-dominated power systems, while the PMR can fulfil this role. We study the type-dependent feature of the PMR and show how PMR can reveal the benefit a GFM inverter can bring to the system compared to a GFL inverter in Section \ref{Section: EMT type-dependent}).

\subsubsection{High SCR while the system is prone to be unstable}\label{Section: EMT high SCR}
We first examine the strength of Bus $4$ (which is electrically equivalent to Bus $2$ due to the negligible impedance of the transformer). The lines' power transfer limits are $X_{14}=1/4$ and $X_{45}=1/1.3$ in per unit, respectively. Both IBRs are grid supporting inverters and generate $1$ p.u. active power. The fault is applied near Bus $4$. 
\begin{figure} [!t]
\centering
\begin{minipage}{0.48\linewidth}
\centerline{\includegraphics[width=5cm]{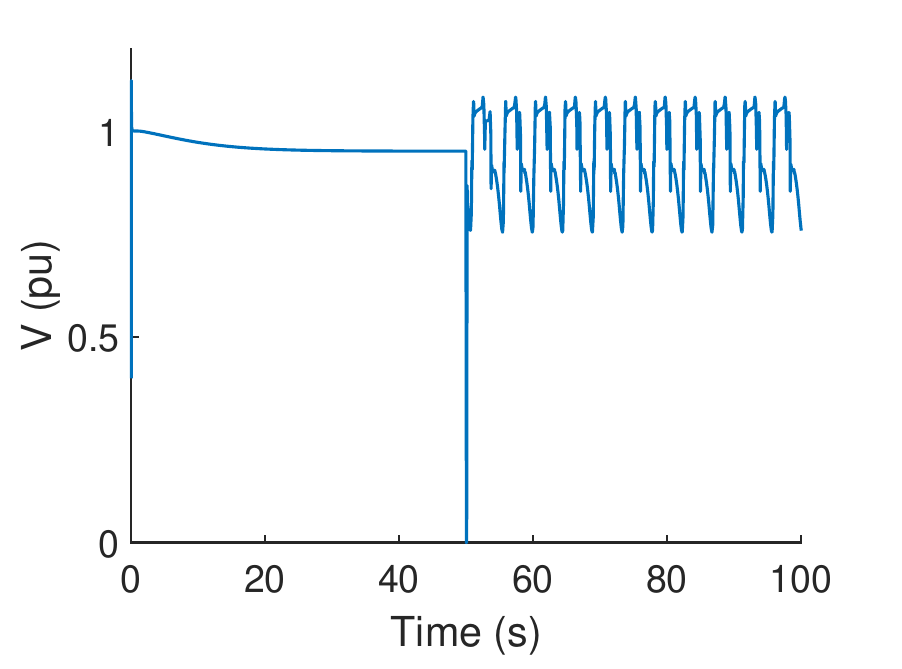}}
\centering
\subfloat{(a)}
\end{minipage}
\hfill
\begin{minipage}{0.48\linewidth}
\centerline{\includegraphics[width=5cm]{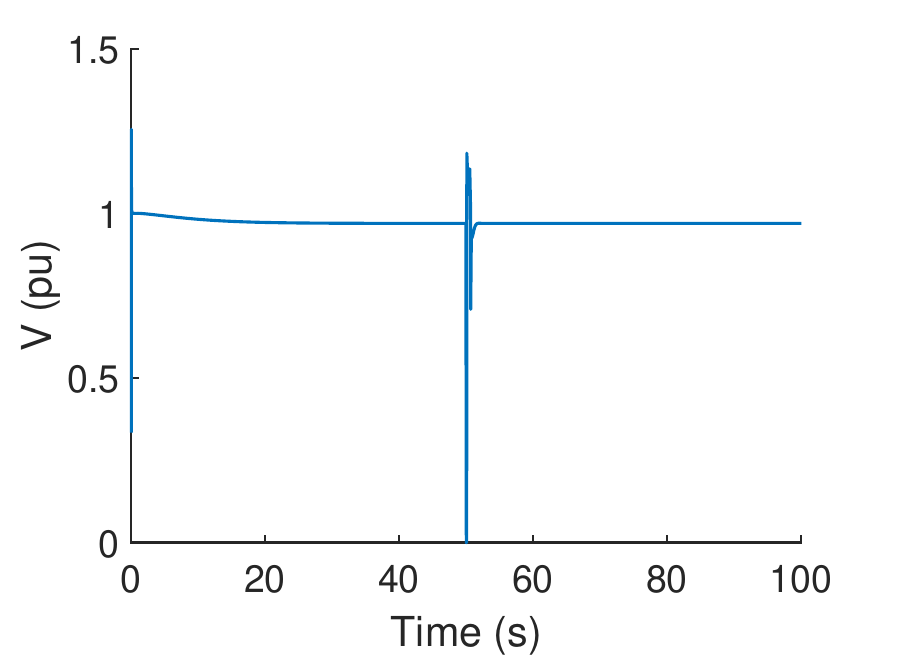}}
\centering
\subfloat{(b)}
\end{minipage}
\vfill
\caption{Voltage at Bus $4$ based on EMT simulations of the two-IBR system (a) without IBR aggregation and (b) with IBR aggregation.} \label{fig: High SCR unstable}
\end{figure}

The voltage at Bus $4$ is presented in Fig. \ref{fig: High SCR unstable} (a). As shown, the system can stably operate at a steady state while it cannot recover from a fault.  The SCR at Bus $4$ is $4$ based on the definition of (\ref{Eqn: SCR pu}), which is considered relatively high. However, the system exhibits oscillatory behavior following the fault. Therefore, ignoring the contribution of IBRs can lead to an inaccurate assessment of the system strength.

Moreover, it should be noted that aggregating the IBRs by simply adding their nominal powers is not a valid solution. At pre-fault steady state, the power transferred from Bus $4$ is $2$ p.u. and $1$ p.u. from Bus $3$, implying that both lines can accommodate such amount of power delivered in a static sense. On the other hand, when we aggregate all IBRs that are located downstream of Bus $4$ so that the equivalent IBR has a nominal power of $2$ p.u., the two-IBR system is transformed into an SPIB system. The EMT simulation results of the equivalent system are shown in Fig. \ref{fig: High SCR unstable} (b). This system can withstand the fault and successfully return to stable operation. The reason is that the strength of Bus $4$ is not merely determined by line $X_{14}$, its downstream line $X_{45}$ plays a crucial role in evaluating the strength of the bus. Ignoring such constraints imposed by other connected lines nearby may lead to an incomplete view of the bus's strength.

\subsubsection{Low SCR while the system is strong}\label{Section: EMT Low SCR}
We then study the strength of Bus $3$ (or Bus $5$ since the equivalent impedance of the transformer is quite small). The lines' power transfer limits are $X_{14}=1/4$ and $X_{45}=1/1.4$ in per unit, respectively. Both IBRs are grid supporting inverters and generate $1$ p.u. of active power. The fault is applied near Bus $5$.
\begin{figure}[!t]
    \centering
    \includegraphics[width=0.9\linewidth]{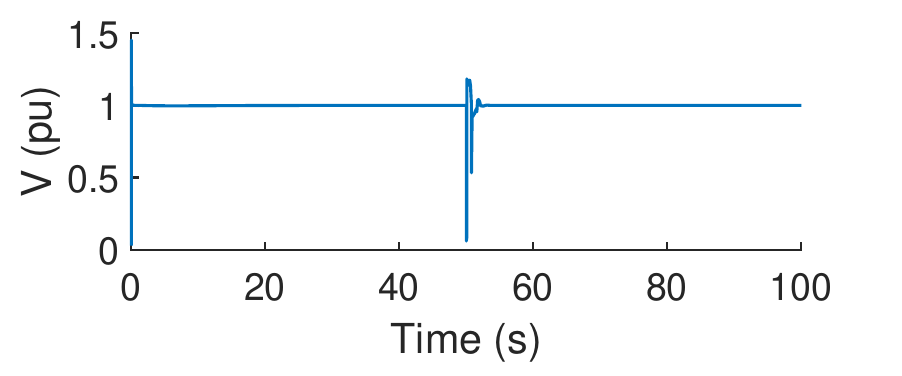}
    \caption{Voltage at Bus $3$ based on EMT simulations of the two-IBR system.}
    \label{fig: Low SCR stable}
\end{figure}

The voltage profile at Bus $3$ is presented in Fig. \ref{fig: Low SCR stable}, where the system is stable after the fault is cleared. Both IBRs are GFL inverters with PV control. Based on the definition in Eqn. (\ref{Eqn: SCR pu}), the SCR of Bus $3$ is $1.037$. It is close to $1$, approaching the system's stable operating boundary. Such a low SCR typically implies a high likelihood of instability when the system experiences a disturbance. However, from Fig. \ref{fig: Low SCR stable}, the system can withstand even a severe three-phase-to-ground fault. To this end, the SCR is no longer a reliable metric to evaluate the system strength for an IBR dominated power system. The system may be strong enough to withstand a severe fault when the SCR is low.

Now we illustrate that the PMR is a better metric for system strength evaluation. We set Bus $2$ and $3$ as PV buses, and Bus $4$ and $5$ as PQ buses with zero active and reactive power. The power flow solution of the steady state is presented in Table \ref{tab: power flow twoIBR system}. Next, we conduct power flow calculations by gradually increasing the active power injection at Bus $3$. The maximum power for Bus $3$ such that the power flow has a converged result is $1.2$ p.u.. Hence, the PMR of this bus is $PMR_3=1.2$. Compared with the conventional SCR, the contribution of IBR at Bus $2$ is involved in the PMR. Precisely, the PMR captures the voltage support function of the IBR at Bus $2$ for which a direct AC voltage control is applied. 
\begin{table}[!t]
    \centering
    \caption{Power flow of the two-IBR system in Fig. \ref{fig: Two IBR system} where $X_{14}=1/4$, $X_{45}=1/1.4$.}
    \begin{tabular}{ccccc}
    \hline
     Bus number & $V_{amp}$ (p.u.) & $\delta$ ($^{\circ}$)& $P$ (p.u.) & $Q$ (p.u.)\\
     \hline
     1 & 1 & 0 & -2 & 0.75\\
     2 & 1 & 34.45 & 1 & 0.92\\
     3 & 1 & 84.73 & 1 & 0.56\\
     4 & 0.95 & 31.54 & 0 & 0\\
     5 & 0.97 & 81.83 & 0 & 0\\
     \hline
    \end{tabular}
    \label{tab: power flow twoIBR system}
\end{table}

\subsubsection{Control type-dependency of the PMR}\label{Section: EMT type-dependent}
The PMR is a type-dependent metric that can distinguish different control modes of the IBR, i.e., GFM, grid supporting, and GFL.

We set the line impedance as $[1/3,1/2]$. The power flow at steady state is listed in Table \ref{tab: power flow twoIBR system2}. To make a comprehensive illustration of the type-dependent feature of the PMR, we fix the system topology and make different combinations in the control mode of those two IBRs. We consider $6$ different combinations of IBRs, i.e., IBR $2$ takes from GFM, grid supporting, and GFL, while IBR $3$ is chosen either grid supporting or GFL. We evaluate the strength of Bus $3$ under each configuration. 
\begin{table}[!t]
    \centering
    \caption{Power flow of the two-IBR system in Fig. \ref{fig: Two IBR system} where $X_{14}=1/3$, $X_{45}=1/2$}
    \begin{tabular}{ccccc}
    \hline
     Bus number & $V_{amp}$ (p.u.) & $\delta$ ($^{\circ}$)& $P$ (p.u.) & $Q$ (p.u.)\\
     \hline
     1 & 1 & 0 & -2 & 0.97\\
     2 & 1 & 47.38 & 1 & 1.00\\
     3 & 1 & 79.82 & 1 & 0.41\\
     4 & 0.95 & 44.46 & 0 & 0\\
     5 & 0.98 & 76.93 & 0 & 0\\
     \hline
    \end{tabular}
    \label{tab: power flow twoIBR system2}
\end{table}

The calculated PMR values of Bus $3$ is listed in Table \ref{tab: PMR Bus 3}. When Bus $3$ is a PQ bus, we assume that the reactive current reference is identical to the one in Table \ref{tab: power flow twoIBR system2}. Among all configurations, the combination of GFM+grid supporting yields the highest PMR, while the lowest PMR appears when both IBRs are GFL. Moreover, comparisons among GFL combinations illustrate that AC voltage control leads to a larger PMR. For this case, voltage control applied to the upstream bus benefits the system. This coincides with the intuition that applying voltage control helps bus voltage become ``stiffer" and improves grid strength.
\begin{table}[!t]
    \centering
    \caption{PMR of BUs $3$ of the two-IBR system with different combinations of IBRs}
    \begin{tabular}{cccc}
    \hline
    Bus $2$ & Bus $3$ & PMR \\
    \hline
     grid supporting & grid supporting & 1.50 \\
     GFL\_PQ & grid supporting & 1.06 \\
     GFM & grid supporting & 1.62 \\
     grid supporting & GFL\_PQ & 1.07 \\
     GFL\_PQ & GFL\_PQ & 1.00 \\
     GFM & GFL\_PQ & 1.09 \\
     \hline
    \end{tabular}
    \label{tab: PMR Bus 3}
\end{table}

\begin{figure} [!t]
\centering
\begin{minipage}{0.48\linewidth}
\centerline{\includegraphics[width=5cm]{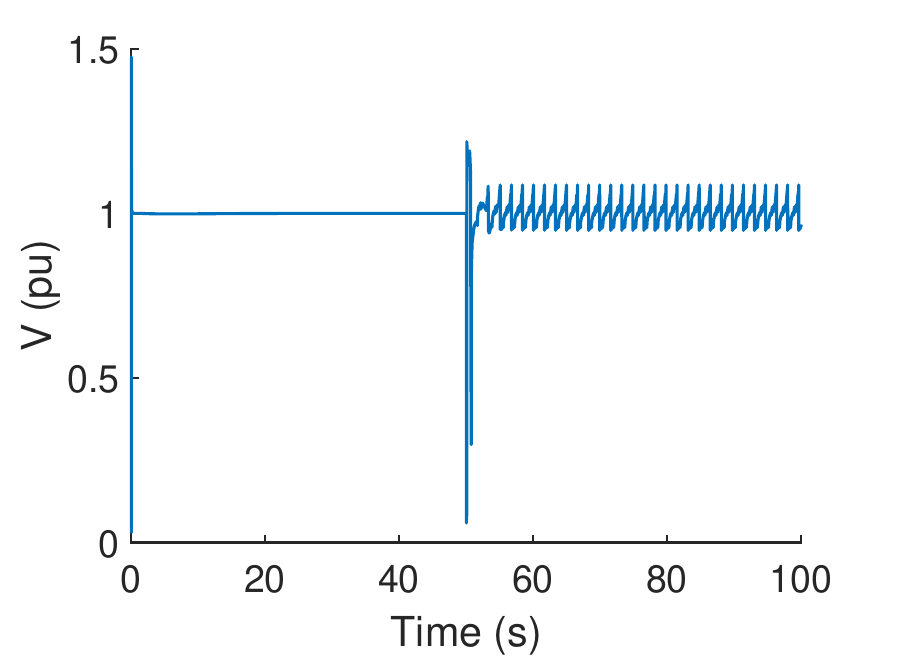}}
\centering
\subfloat{(a)}
\end{minipage}
\hfill
\begin{minipage}{0.48\linewidth}
\centerline{\includegraphics[width=5cm]{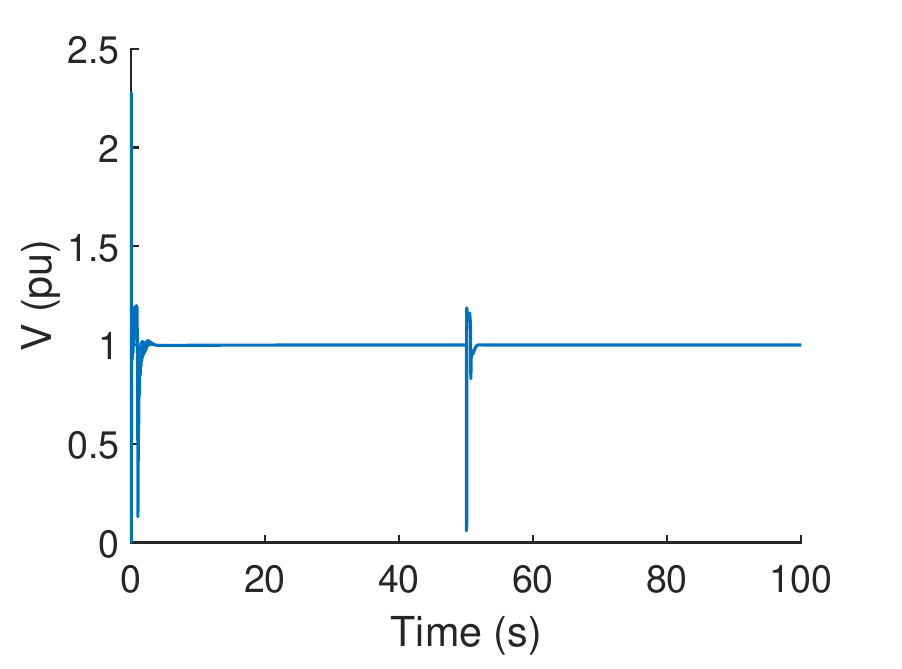}}
\centering
\subfloat{(b)}
\end{minipage}
\vfill
\caption{Voltage at Bus $3$ based on EMT simulations of the two-IBR system (a) grid supporting and (b) GFM.} \label{fig: type dependent 1.05}
\end{figure}

We also verify the stability by conducting EMT simulations, and the voltage at Bus $3$ is presented in Fig. \ref{fig: type dependent 1.05}. We omit the voltage profile of Bus $2$ since it shows similar results. The power output of the IBR $3$ is set to $1.05$ p.u. In grid supporting mode, the system oscillates after the fault is cleared. On the contrary, connecting a GFM at Bus $2$ facilitates the recovery of the system after a fault. The GFM behaves like a stiff voltage source which provides strong voltage support to the grid. Comparing the first three rows of Table \ref{tab: PMR Bus 3}, the PMR for the GFM case is the largest, which is in line with the EMT simulation results. \textit{In conclusion, in terms of grid supporting, GFM$>$grid supporting$>$GFL.}

The calculated PMR values of Bus $2$ is listed in Table \ref{tab: PMR Bus 2}. When Bus $2$ is a PQ bus, we assume that the reactive current reference is identical to the one in Table \ref{tab: power flow twoIBR system2}. Among all configurations, the combination of GFM+grid supporting yields the largest PMR while the smallest PMR appears when both IBRs are GFLs. Similarly, the bus with voltage support has higher strength than that without voltage controls. GFM inverters can significantly enhance the system strength.
\begin{table}[!t]
    \centering
    \caption{PMR of BUs $2$ of the two-IBR system with different combinations of IBRs}
    \begin{tabular}{cccc}
    \hline
    Bus $2$ & Bus $3$ & PMR \\
    \hline
     grid supporting & grid supporting & 1.61 \\
     grid supporting & GFL\_PQ & 1.46 \\
     grid supporting & GFM &  3.00\\
     GFL\_PQ & grid supporting & 1.10 \\
     GFL\_PQ & GFL\_PQ &  1.00\\
     GFL\_PQ & GFM &  1.81\\
     \hline
    \end{tabular}
    \label{tab: PMR Bus 2}
\end{table}

\subsubsection{PMR and the ROA}
\begin{figure}[!t]
  \centering
  \includegraphics[width=2.8 in]{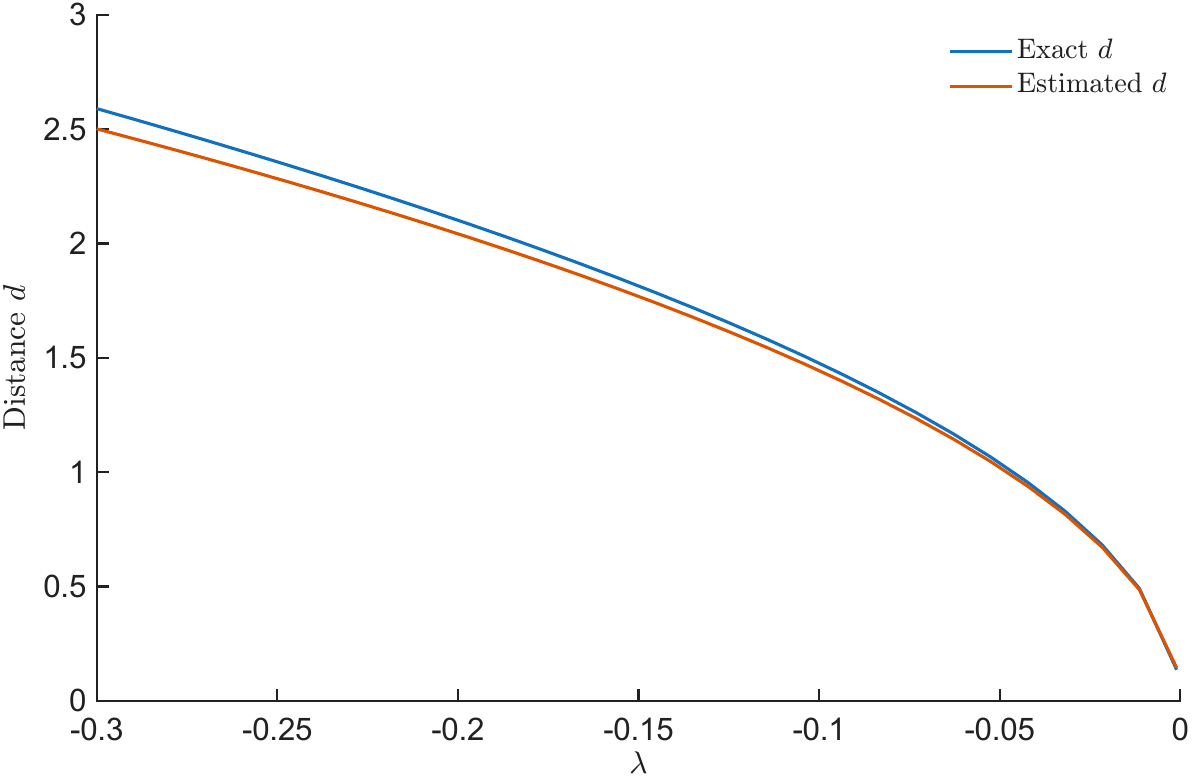}\\
  \caption{Illustration of the estimated distance and exact distance curves.}\label{fig: Distance lambda}
\end{figure}

We now illustrate the relationship between PMR and the distance $d=||SEP-UEP||$ by numerical studies. Consider the case shown as the first row in Table \ref{tab: PMR Bus 2}, where the PMR of Bus $2$ is $1.61$. Given a $\lambda$, We obtain the exact distance by numerically solving nonlinear power flow equations to identify the UEP, while the estimated distance is calculated based on (\ref{Eqn: Estimated distance}). Fig. \ref{fig: Distance lambda} presents the variation of the exact (blue curve) and estimated (red curve) distance for $\lambda\in [-0.3, 0)$. As $\lambda\rightarrow 0$, both the exact and the estimated ones reduce, and the difference between them decreases as well. The reason can be explained based on the derivation in Section \ref{Section PMR ROA}, i.e., when $|\lambda|$ increases, the higher-order terms in Taylor's expansion have an inignorable impact on the distance $d$.

\subsubsection{PMR and transient stability}
\begin{figure}[!t] 
\centering
\begin{minipage}{0.48\linewidth}
\centerline{\includegraphics[width=4.5cm]{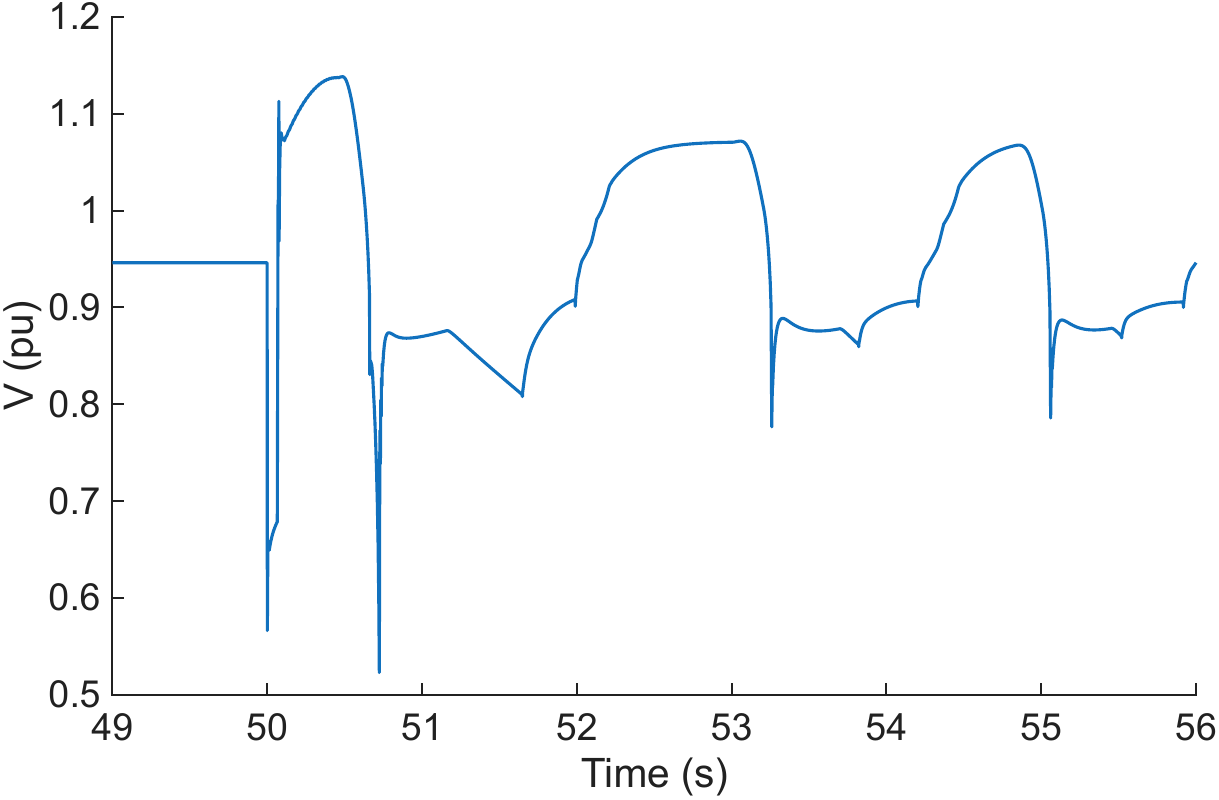}}
\centering
\subfloat{(a)}
\end{minipage}
\hfill
\begin{minipage}{0.48\linewidth}
\centerline{\includegraphics[width=4.5cm]{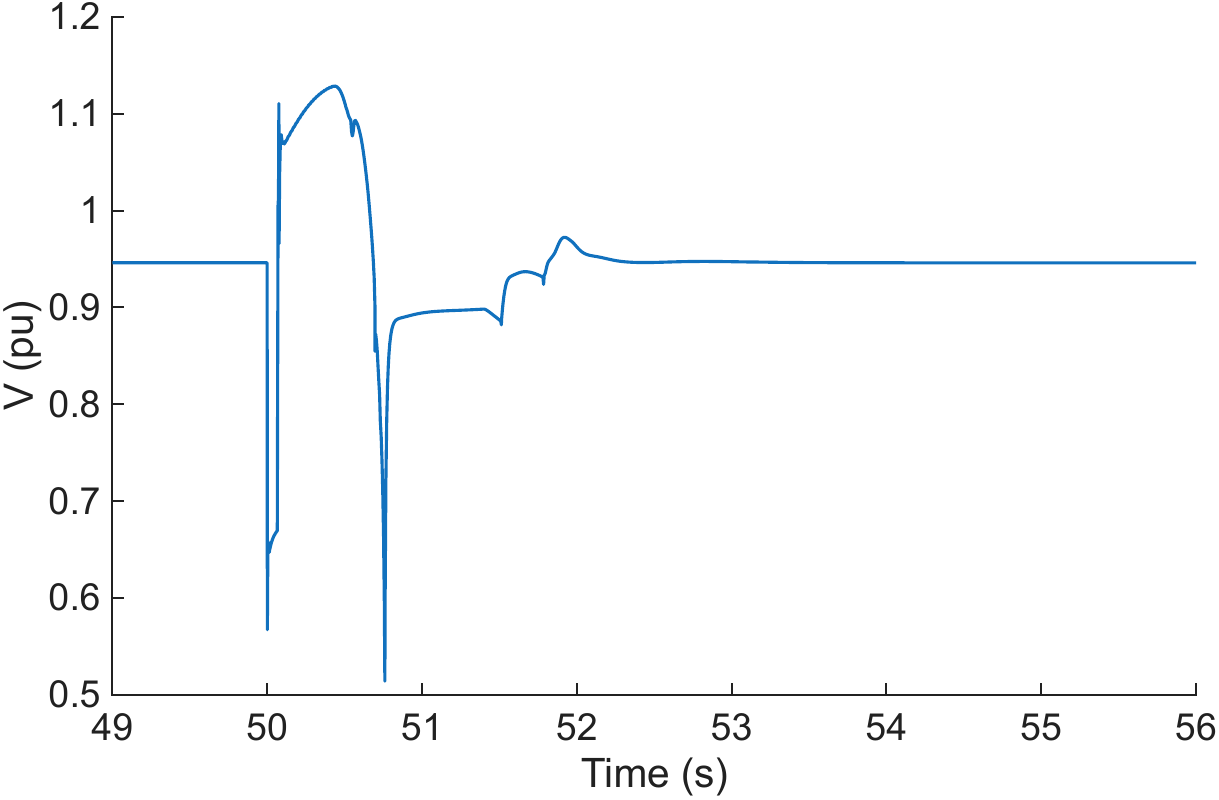}}
\centering
\subfloat{(b)}
\end{minipage}
\caption{Unstable (a) and stable (b) case: terminal voltages of the grid supporting IBRs at Bus $2$.} \label{fig: discussion V bus 2}
\end{figure}

It is worth nothing that the proposed PMR is served as a metric for large-signal system strength assessment instead of a rigorous transient stability certificate to conclude whether the system is stable or not in a post-fault stage. In fact, the instability is closely related to the interaction of the control loops of the IBR. To illustrate this, we consider the two-IBR system with line impedance $X_{14}=1/3$ and $X_{45}=1/2$. The nominal power of IBR at Bus $2$ is set to $1$ p.u. and $1.05$ p.u. for the one at Bus $3$. Both IBRs works in grid supporting mode. As shown in Fig. \ref{fig: discussion V bus 2} (a), the system cannot recover to its normal operation after clearing the fault. The terminal voltage and the PLL angle of the IBR at Bus $2$ are oscillating. Nevertheless, if we reduce the bandwidth of the PLL to a half, the system is brought back to normal operation, as shown in Fig. \ref{fig: discussion V bus 2} (b). The stable and unstable cases have the same PMR while the actual transient stability is different. 

In PMR calculation, we tend to use static information to reveal the transient behaviour of the system. For instance, we assume the GFM inverter is a $V\theta$ bus, although the angle of the GFM is actually not constant during the transient period; the grid supporting inverter is assumed as a PV bus, whereas the voltage is not constant during the transient period. These assumptions make the PMR not a rigorous characterization of the stability of the power system. However, PMR captures the key features of distinct control modes of IBRs. Exploiting such features, the degree of contribution to the grid strength from different types of IBRs is involved.

\section{Conclusion}
A metric, PMR, for large-signal grid strength assessment of IBR-dominated power systems is proposed in this paper. It is power flow calculation-based, control type-dependent metric that involves the contributions of IBRs. PMR achieves a whole-system assessment that evaluates the impacts of multiple infeeds which share the power transfer capacity of a transmission system. The theoretical foundation of PMR is thoroughly explored by studying the relationship between PMR and ROA. Comprehensive case studies have validated that although PMR is not a rigorous transient stability certificate, it can reveal the system strength from a static perspective. That is, a higher PMR (of a concerned bus) implies a stronger connecting point.
\ifCLASSOPTIONcaptionsoff
  \newpage
\fi

\bibliographystyle{IEEEtran}
\bibliography{refs.bib}

\end{document}